\documentclass[aip,twocolumn,pra,showpacs]{revtex4}

\usepackage{epsfig}
\usepackage{graphics}
\usepackage{isolatin1}
\usepackage{amssymb,amsmath}

\newcommand{\HC}{H_{\mathrm{c}}}             
\newcommand{\eins}{1\!\! 1}                  
\newcommand{\e}{\mathrm{e}}                  
\newcommand{\ket}[1]{|#1\rangle}             
\newcommand{\bra}[1]{\langle #1|}            
\renewcommand{\v}[1]{\mathbf{#1}}            
\renewcommand{\Im}{\mathrm{Im}}              
\newcommand{\mt}[1]{\mathrm{#1}}             

\begin{document}

\title{Quantum arrival times and operator normalization}

\author{Gerhard C. Hegerfeldt}
\author{Dirk Seidel}
\affiliation{Institut für Theoretische Physik, Universität Göttingen,
  Bunsenstrasse 9, 37073 Göttingen, Germany}
\author{J. Gonzalo Muga}
\affiliation{Departamento de Qu\'{\i}mica-F\'{\i}sica, Universidad del
Pa\'{\i}s Vasco, Apartado Postal 644, 48080 Bilbao, Spain}
   
\begin{abstract}
A recent approach to arrival times used the fluorescence 
of an atom  entering a laser illuminated region and the resulting  
arrival-time distribution was close to the axiomatic distribution of
Kijowski, but not exactly equal, neither in limiting cases nor after
compensation of reflection losses by normalization on the level of
expectation values. In this paper we employ a normalization on the
level of operators, recently proposed in a slightly different
context. We show that in this case  the
axiomatic arrival time distribution of Kijowski is recovered as a
limiting case. In
addition, it is shown that Allcock's complex potential model is also a
limit of the physically motivated fluorescence approach and
connected to Kijowski's distribution through operator normalization.
\end{abstract}
\maketitle

\section{Introduction}
The quantum mechanical analog of the arrival time of a particle at a
given location is physically very interesting, and for wave packets
which are spreading in space this is a highly nontrivial subject. It
is a particular case of the description of    
time observables in quantum mechanics, i.e., times as random instants 
-- such as the arrival times -- or durations, e.g. dwell or sojourn
times. For recent 
reviews cf. Refs. \cite{ML00,tqm}. Difficulties in the formulation of
quantum arrival times and attempts to overcome these were presented
e.g. in Refs.
\cite{Allcock,Kijowski,Werner,Yamada,Mielnik94,Leavens,BJ,GRT,Halliwell,
Finkelstein,Toller,KW,AOPRU98,
Aharonov-PR-1961,Leon-PRA-2000,BSPME00,BEM01a,Baute01,DM,AHN99,Galapon,
Wlodarz,Leavens02,
EMNR03,Damborenea-PRA-2002,Damborenea-JPB-2003}.
In particular, the lack of a self-adjoint arrival-time operator
conjugate to the free Hamiltonian lies at the core of these difficulties.
 
Allcock \cite{Allcock} modeled a simplified detection procedure in the
region $x>0$ by means of a complex absorption potential. Because of
reflection, he disregarded strong absorption and only considered the
weak absorption limit, in which the detection takes a long time but
all particles are eventually detected. Under the assumption that the measured
arrival-time distribution was a convolution of an ideal distribution
and an apparatus function he suggested for the unknown ideal
distribution an {\it approximate} positive expression, obtaining
the (not semidefinite positive) current
density as the exact solution. 
Somewhat pessimistically
he argued that a fully satisfactory, apparatus independent,
arrival-time distribution could not be defined.

In contrast, Kijowski \cite{Kijowski} (cf. also Ref. \cite{Werner})
pursued an axiomatic approach modeled on the classical case and
obtained as arrival-time distribution at $x=0$ for a free particle
of mass $m$ coming in from the left with initial state
$\tilde{\psi}(k)$ ($k$ is the wavenumber) 
an expression which, in the one-dimensional
case, is given by
\begin{equation} \label{Kijowski}
\Pi_K(t)= \frac{\hbar}{2\pi m}~ {\Big |}\! \int  dk
 ~ \tilde{\psi}(k) \sqrt{k} ~ \e^{-i\hbar k^2t/2m} {\Big |}^2.
\end{equation}
Surprisingly, this coincides with the approximate expression suggested
by Allcock \cite{Allcock}.

Much more recently, the distribution $\Pi_K$ has been related to the 
positive operator valued measure (POVM) generated by the 
eigenstates of the Aharonov-Bohm (maximally 
symmetric) time-of-arrival operator \cite{Aharonov-PR-1961,ML00,Giannitrapani-IJTP-1997}, and
this method 
emphasizes the fact that self-adjointness is not necessary
to generate quantum probability distributions. The distribution  
has  also been generalized 
for the case where the particle is affected by interaction
potentials \cite{Leon-PRA-2000,BSPME00, BEM01a}
and for multi-particle systems
\cite{Baute01}.   

Yet, the status of Kijowski's distribution has remained unclear and 
controversial \cite{ML00,Leavens02,EMNR03}. 
As an ideal
distribution, some of its properties or of its generalizations 
have been questioned \cite{Leavens02}
or considered 
to be puzzling \cite{ML00},
and its ``operational'' interpretation, apart from the approximate 
connection found by Allcock,  
has remained elusive \cite{Damborenea-PRA-2002}. 

In two recent papers \cite{Damborenea-PRA-2002,Damborenea-JPB-2003}, a
procedure to determine arrival times of 
quantum mechanical particles has been discussed,
which is based on 
the detection of fluorescence photons emitted when
a two-level atom enters a laser-illuminated region.
In general, due to partial reflection of
the atoms by the laser field, not all of them emit photons and
hence some go undetected. Therefore the measured distribution of arrival
times is not  normalized to one. To normalize  the distribution,
division by its time integral was considered
(`normalization on the level of 
expectation values'). In some cases this gave good agreement with the
axiomatically proposed distribution of Kijowski \cite{Kijowski}, 
and parameter regimes where this agreement could be found were 
described. Analogously to Allcock´s absorption model, the current density 
could be obtained exactly in the weak laser driving limit by 
deconvolution, and 
strong driving was problematic because of the atomic reflection. 
The coincidence between the results of the simplified complex potential
model and the more realistic and detailed laser-atom model is not
accidental and will be explained below. 
  
Also recently, Brunetti and Fredenhagen \cite{Brunetti-PRA-2002} have 
proposed a general
construction of an observable measuring the `time of occurrence' of some 
event. 
This construction involved a  unitary time development and
a normalization procedure on the level of operators, not on the level of
expectation values. For this purpose 
they constructed  a positive operator on the orthogonal 
complement of the states on which the time of occurrence
is infinite or zero and used its square root for normalization.
This normalization procedure was in particular applied to  sojourn or
dwell times.

In this paper it will be shown that normalization on the level of operators
can also be applied to the approach to arrival times of
Ref.~\cite{Damborenea-PRA-2002} which
uses spontaneous photon emissions and,
as a technical device, a `conditional' non-unitary time-development.
As a  result we obtain quite simple and explicit expressions. In 
particular, the physically attractive limit of strong laser field and
fast spontaneous emission can be performed and shown to exactly yield the
axiomatic distribution of  Kijowski \cite{Kijowski}.

In the next section we briefly review the results of
Refs.~\cite{Damborenea-PRA-2002} and \cite{Damborenea-JPB-2003} and
then calculate the operator normalized arrival-time distribution.
In Sections \ref{III} and \ref{IV}, fast spontaneous emission and
strong laser fields are considered in different limits. 
Finally, a connection between the fluorescence approach and complex
absorption models is exhibited.

\section{operator-normalized arrival times}

In the one-dimensional model of Ref. \cite{Damborenea-PRA-2002}, a
two-level atom wave packet impinges on a perpendicular  
laser beam at resonance with the atomic transition. Using the
quantum jump approach  \cite{Hegerfeldt93} the continuous
measurement of the fluorescence 
photons is simulated by a repeated projection onto no-photon or 
one-photon subspace every $\delta t$, a time interval large enough
to avoid the Zeno effect, but smaller than any other characteristic time.
The amplitude for the
undetected atoms in the interaction picture for the internal Hamiltonian 
obeys, 
in a time scale coarser than $\delta t$, and using the
rotating wave and dipole 
approximations, an effective Schr\"odinger equation
governed by the complex ``conditional''  Hamiltonian
(the hat is used to distinguish momentum and position operators from 
the corresponding c-numbers)
\begin{equation}
  \HC = \frac{\hat{p}^2}{2m} -i\hbar\frac{\gamma}{2}\ket{2}\bra{2} +
  \frac{\hbar\Omega}{2}\Theta(\hat{x}) 
\left(\ket{2}\bra{1} + \ket{1}\bra{2}\right), 
\end{equation}
where the 
ground state $|1\rangle$ is in vector-component notation ${1 \choose 0}$,
the excited state $|2\rangle$ is ${0 \choose 1}$, $\Theta(x)$ is the
step function, and $\Omega$ is the Rabi frequency, which gives  the
interaction strength with the laser field. 
 
To obtain the time development under $H_{\rm
c}$ of a wave packet incident from the left 
one first solves the stationary equation
\begin{equation}\label{eigenvalue}
H_{\rm  c}{\bf \Phi}_k = E_k {\bf \Phi}_k,~~~~~{\rm where}~~{\bf
  \Phi}_k(x)\equiv{\phi_k^{(1)}(x)\choose\phi_k^{(2)}(x)} 
\end{equation}
for scattering states 
with real energy 
$$
E_k = \hbar^2k^2/2m 
$$
which are incident from the left ($k>0$). These are given by
\cite{Damborenea-PRA-2002}
\begin{equation}
  \v{\Phi}_k(x) = \frac{1}{\sqrt{2\pi}} \left\{ 
    \begin{array}{l}
      \displaystyle{ {\e^{i kx}+R_1\,\e^{-i kx}} \choose 
    {R_2\,\e^{-i qx}}},~~~~~ x\leq 0 \\
      C_{+}\ket{\lambda_{+}}\e^{i k_{+}x} +
      C_{-}\ket{\lambda_{-}}\e^{i k_{-}x},\\ \qquad\qquad\qquad\qquad
      \qquad x\geq 0
    \end{array} \right.
\end{equation}
where
\begin{eqnarray}
  q &=& \sqrt{k^2+i m\gamma/\hbar}\\
  k_{\pm} &=& \sqrt{k^2 - 2m\lambda_{\pm}/\hbar}
\end{eqnarray}
with $\Im\,q >0$, $\Im\, k_{\pm}>0$, and where
\begin{eqnarray}
  \lambda_{\pm} &=& (-i\gamma \pm i\sqrt{\gamma^2 - 4\Omega^2})/4\\
  \ket{\lambda_{\pm}} &=& 1 \choose {2\lambda_{\pm}/\Omega}
\end{eqnarray}
are eigenvalues and eigenvectors of the matrix
$\frac{1}{2}$${0\quad\Omega}\choose{\Omega\,\,\, -i\gamma}$.
The coefficents $R_1,R_2, C_{+}, C_{-}$
follow from the matching conditions
at $x=0$ as
\begin{eqnarray}
  R_1 &=& [\lambda_{+}(q+k_{+})(k-k_{-})\nonumber\\ &&-
  \lambda_{-}(q+k_{-})(k-k_{+})]/D\\
  R_2 &=& k(k_{-}-k_{+})\Omega/D\\
  C_{+} &=& -2k(q+k_{-})\lambda_{-}/D\\
  C_{-} &=& 2k(q+k_{+})\lambda_{+}/D
\end{eqnarray}
with the common denominator
\begin{equation}
D = \lambda_{+}(q+k_{+})(k+k_{-}) - \lambda_{-}(q+k_{-})(k+k_{+}).
\end{equation}
By decomposing an initial state as a superposition of eigenfunctions,
one obtains its conditional time development. This is easy for an initial 
ground-state wave packet
coming in from the far left side 
in the remote past.   Indeed, 
\begin{eqnarray}\label{2.9}
{\bf \Psi}(x,t)= \int_0^\infty dk \,\widetilde{\psi}(k) \,{\bf \Phi}_k 
(x)\,e^{-i \hbar k^2 t/2m}
\end{eqnarray} 
describes the conditional time development of a state which in the
remote past behaves like a wave packet in the ground-state coming in
from the left, with
$\widetilde{\psi}(k),~k>0$,
the momentum amplitude it would have at $t=0$ as a
freely moving packet.
The probability, $N_t$, of no photon detection 
up to time $t$ is given by \cite{Hegerfeldt93}
\begin{eqnarray} \label{eq:Nt}
  N(t) &=& ||{\bf \Psi}_t||^2, 
\end{eqnarray} 
and the probability density, $\Pi (t)$, for the first photon detection is
given by
\begin{eqnarray} \label{eq:Pi}
\Pi(t) &=& -\frac{d N(t)}{d t}. 
\end{eqnarray}
For the   two-level system under consideration one has $H_{\rm c} -
H_{\rm c}^\dagger = - i\gamma\hbar |2\rangle\langle2|$, and thus 
\begin{eqnarray} \label{eq:Pi2}
\Pi(t) = \gamma ||\psi^{(2)}_t ||^2.
\end{eqnarray}
The integral of the distribution $\Pi(t)$ is in general smaller than
1, in fact
\begin{equation} \label{integral}
\int_{-\infty}^{\infty} d t~ {\Pi}(t) = 1 - N(\infty)
\end{equation}
and this was used in Ref. \cite{Damborenea-PRA-2002} for normalization
on the level of expectation values.

In order to employ operator normalization we rewrite Eq. (\ref{eq:Pi})
in operator form, and to do this we go to the interaction picture with
respect to $H_0 = \hat{p} ^2/2m$, 
\begin{eqnarray} \label{interaction}
  \HC^I &=& \e^{i H_0 t/\hbar}(\HC- H_0) \e^{-i H_0 t/\hbar}\nonumber\\
U^I_{\rm c}(t,t_0) &=& \e^{i H_0 t/\hbar}\e^{-i \HC (t- t_0)/\hbar}\e^{-i H_0
  t_0/\hbar},\nonumber\\&&  
\end{eqnarray}
where $U^I_{\rm c}$ is the conditional time development  corresponding
to $\HC^I$. Then Eq. (\ref{2.9}) can be written as 
\begin{eqnarray} \label{2.9a}
{\bf \Psi}_t =\e^{-i H_0t/\hbar}\,U^I_{\rm c}(t,-\infty)\ket{\psi}\ket{1}, 
\end{eqnarray}
and Eq. (\ref{eq:Nt}) as 
\begin{eqnarray} \label{2.10}
N(t) = \bra{1}\bra{\psi}\hat{N}_t\ket{\psi}\ket{1},
\end{eqnarray}
with
\begin{eqnarray} \label{2.11}
\hat{N}_t = U^I_{\rm c}(t,-\infty)^{\dagger}\,U^I_{\rm c}(t,-\infty).
\end{eqnarray}
Similarly,
\begin{eqnarray} \label{2.12}
\Pi(t) = \bra{1}\bra{\psi}\hat{\Pi}_t\ket{\psi}\ket{1},
\end{eqnarray}
with
\begin{eqnarray} \label{2.13}
\hat{\Pi}_t &=&- \frac{d \hat{N}(t)}{d t}\\ &=& \gamma \,
U^I_{\rm c}(t,-\infty)^{\dagger}\ket{2}\bra{2}U^I_{\rm c}(t,-\infty). 
\end{eqnarray}
In analogy to Eq. (\ref{integral}) we consider the integral 
\begin{eqnarray} \label{2.13a}
 \int_{-\infty}^{\infty} d t~ \hat{\Pi}_t 
= \eins -
  \hat{N}_{\infty} 
\end{eqnarray} 
and define the operator $\hat{B}$  on the incoming states (with internal
ground state) through  its matrix elements as
\begin{eqnarray}\label{eq:def_B}
\bra{1}\bra{\varphi}\hat{B}\ket{\psi}\ket{1} 
= \bra{1}\bra{\varphi}\eins -
  \hat{N}_{\infty} \ket{\psi}\ket{1}.
\end{eqnarray} 
The operator $\hat{B}$ can be easily calculated as follows. From 
Eq. (\ref{2.9}) one sees that for large $t$ the second component of
${\bf \Psi}(x,t)$ is damped away and therefore only the
reflected wave remains,
\begin{equation}\label{2.9b}
{\bf \Psi}(x,t) \simeq
 \int_0^\infty dk \,\widetilde{\psi}(k)R_1(k) \e^{ikx}
\,\e^{-i \hbar k^2 t/2m}\ket{1}.
\end{equation} 
for large $t$.
Pulling $\e^{-i \hbar k^2 t/2m}$ out from the integral as $\e^{-i
H_0t/\hbar}$ one sees, from Eqs. (\ref{2.9a}) and (\ref{2.11}), that  
\begin{equation}\label{2.14}
U_{\rm c}(\infty,-\infty)\ket{\psi}\ket{1} = \int_0^\infty dk
\,\widetilde{\psi}(k)R_1(k)\ket{-k} \ket{1}.
\end{equation} 
Taking the scalar product with $U_{\rm
c}(\infty,-\infty)\ket{\varphi}\ket{1}$  one finds from
Eq. (\ref{eq:def_B}), and in $k$ space, 
\begin{equation}\label{2.15}
  \bra{1}\bra{k}\hat{B}\ket{k'}\ket{1} =
  \Bigl(1-\overline{R_1(k)}R_1(k')\Bigr)\delta(k-k').
\end{equation} 

Hence, on the incoming states, one can define the operator 
\begin{eqnarray}\label{eq:Pi_ON_op}
\hat{\Pi}^{\text{\tiny ON}}_t = \hat{B}^{- 1/2} \hat{\Pi}_t \hat{B}^{- 1/2}.
\end{eqnarray} 
From Eqs. (\ref{eq:def_B}) and (\ref{2.13}) one sees that
$\int^{\infty}_{-\infty} dt\,\hat{\Pi}^{\text{\tiny ON}}_t= \eins$ and so the
probability distribution
\begin{eqnarray}\label{2.16}
  \Pi^{\text{\tiny ON}}(t) \equiv \bra{1}\bra{\psi}\hat{\Pi}^{\text{\tiny ON}}_t\ket{\psi}\ket{1}
\end{eqnarray} 
is normalized to 1. From Eqs. (\ref{2.13}) and (\ref{2.15}) one
finally obtains
\begin{widetext}
\begin{eqnarray} \label{eq:Pi_ON_fin}
 \Pi^{\text{\tiny ON}}(t) &=& \gamma \int_{-\infty}^{\infty} d x~\int d kd k'~ 
  \overline{\tilde{\psi}(k)} \tilde{\psi}(k')\,(1-|R_1(k)|^2)^{-1/2}
  \,(1-|R_1(k')|^2)^{-1/2} \nonumber\\
  &&\qquad\qquad\qquad\times\,\,\e^{i\hbar (k^2-k'^2)t/2m}
  \overline{\phi^{(2)}_k(x)} \phi^{(2)}_{k'}(x).
\end{eqnarray}
\end{widetext}
Since $|R_1(k)| < 1$, $\hat{B}$ is not only a positive operator but
also its inverse square-root exists. 

\begin{figure}[htbp]
  \begin{center}
    \includegraphics[width=7cm]{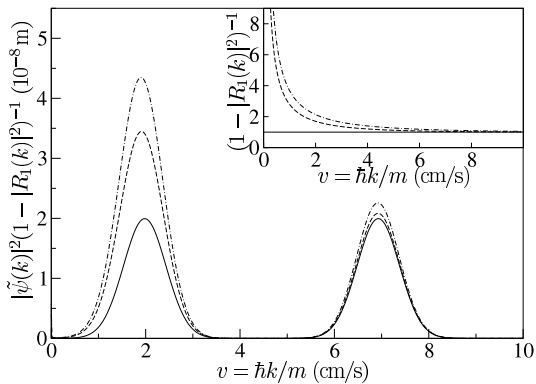}
    \caption{Operator normalization viewed as change of initial
      momentum distribution. Two Gaussian momentum wave packets
      with $\langle v\rangle_1 = 2\,\mt{cm/s}$, $\langle v\rangle_2
      = 7\,\mt{cm/s}$, $\Delta v_1 = \Delta v_2 = 0.48\,\mt{cm/s}$,
      without (solid line) and with operator normalization
      for $\Omega = 0.66\gamma$ (dashed line) and $\Omega = \gamma$
      (dot-dashed line). All figures are for the transition $6^2
      P_{3/2} - 6^2 S_{1/2}$ of cesium with $\gamma =
      33.3\cdot 10^6\,\mt{s}^{-1}$.} 
    \label{fig:ON_infl}
  \end{center}
\end{figure}

Operator normalization can be viewed as a
change in the incident momentum distribution $\tilde{\psi}(k)$ by a
factor of $(1-|R_1(k)|^2)^{-1/2}$. The effect of this factor on a
Gaussian wave packet is shown in Fig.~\ref{fig:ON_infl}.
For mean
initial velocities of the order of $\mt{cm/s}$ a single wave packet is
multiplied  by a nearly constant factor. Only for very slow particles and
$\Omega\gg\gamma$ a distortion of the packet occurs. In this region
the amplification of the slow components by operator normalization
leads to an additional delay of $\Pi^{\text{\tiny ON}}(t)$ compared to
$\Pi(t)$.   

\section{The axiomatic arrival-time distribution as a limit} \label{III}

In this section it will be shown that the operator-normalized
distribution $\Pi^{\text{\tiny ON}}(t)$ approaches Kijowski's axiomatic
distribution for large $\gamma$ and $\Omega$, with $\gamma^2/\Omega^2
= $ const. We put $\alpha \equiv\sqrt{1-4\Omega^2/\gamma^2}$ and find,
for large $\gamma$, 
\begin{eqnarray}\label{3.1}
  \lambda_{\pm} &=& \frac{i\gamma}{4}(-1\pm\alpha)\nonumber\\
  q &\simeq& \sqrt{\frac{i m\gamma}{\hbar}}\nonumber\\
  k_{\pm} &\simeq & q\sqrt{\frac{1\mp\alpha}{2}}\nonumber\\
  R_1 &\simeq & -1-\frac{2i
    k}{\gamma^{1/2}}\sqrt{\frac{i\hbar}{m}}C_1(\alpha)\nonumber\\
  R_2 &\simeq& -\frac{k}{\gamma^{1/2}}\sqrt{\frac{i\hbar}{m}}C_2(\alpha),
\end{eqnarray}
to leading order in $\gamma$, where the constants $C_i$  are given
explicitly in the Appendix. From this one obtains
\begin{eqnarray}
  (1-|R_1(k)|^2)^{-1/2}\,(1-|R_1(k')|^2)^{-1/2}\nonumber\\ \simeq
  \frac{1}{4C_1}\sqrt{\frac{2m\gamma}{\hbar kk'}}
\end{eqnarray}
and
\begin{multline}\label{3.3}
  \gamma \overline{\Phi^{(2)}_k(x)} \Phi^{(2)}_{k'}(x)\\ \simeq
  \frac{\hbar kk'}{2\pi m}\Bigl\{
      \Theta(-x) C_2^2 \e^{-i(q-\overline{q})x} \\
      +  \Theta(x) \frac{16}{C_3^2}\frac{\Omega^2}{\gamma^2} 
      \Big|\Bigl(1+\sqrt{\frac{1+\alpha}{2}}\Bigr) \e^{i k_{+} x}\\ -
      \Bigl(1+\sqrt{\frac{1-\alpha}{2}}\Bigr) \e^{i k_{-} x}\Big|^2 \Bigr\}.
\end{multline}
Then $\Pi^{\text{\tiny ON}}(t)$ becomes, for large $\gamma$ and
$\Omega^2/\gamma^2 = $ const,   
\begin{widetext}
\begin{eqnarray} \label{eq:Pi_ON_lim1}
\!\!\!\!\!\!\!\!  \Pi^{\text{\tiny ON}}(t) &\simeq& \frac{\hbar}{2\pi m} \int
d kd k'~ 
  \overline{\tilde{\psi}(k)} \tilde{\psi}(k') 
  \e^{i\hbar (k^2-k'^2)t/2m} \sqrt{kk'} \nonumber\\
&&  \times \frac{1}{4C_1}\sqrt{\frac{2m\gamma}{\hbar}}
  \int_{-\infty}^{\infty} d x~  \Big\{ \Theta(-x)
      C_2^2 \e^{-i(q-\overline{q})x}\nonumber\\
&&~~~~~ + \Theta(x)
      \frac{16}{C_3^2}\frac{\Omega^2}{\gamma^2} 
      \Big|\Bigl(1+\sqrt{\frac{1+\alpha}{2}}\Bigr) \e^{i k_{+} x} -
      \Bigl(1+\sqrt{\frac{1-\alpha}{2}}\Bigr) \e^{i k_{-} x}\Big|^2
      \Big\}. 
\end{eqnarray}
\end{widetext}
Inserting $q$ and $k_{\pm}$ from (\ref{3.1}) one sees that the expression
after $\times$ is independent of $k$ and $k'$. One can insert $C_i$
from the Appendix A and explicitly calculate the integral over $x$,
but it is
easier to note that the term before  $\times$ is just
Kijowski's distribution, which is normalized to 1, and therefore the
the expression after $\times$ must equal 1. 

\begin{figure}[htbp]
  \begin{center}
    \includegraphics[width=7cm]{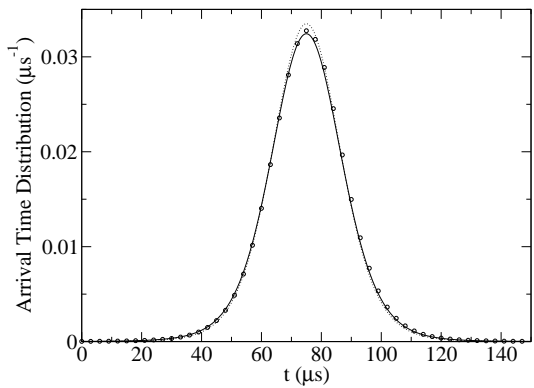}
    \caption{Good agreement of $\Pi^{\text{\tiny ON}}$ (circles) 
      with $\Pi_K$ (solid line) for large but finite $\gamma$,
      $\gamma = 10\gamma_{\mt{Cesium}}$, $\Omega = 0.33\gamma$. The
      initial Gaussian wave packet is chosen to become minimal when
      its center arrives at $x=0$ (in the absence of the laser) to
      enhance the difference between $\Pi_{K}$ and the flux
      (dotted line); $\langle
      v\rangle = 0.9\,\mt{cm/s}$, $\Delta x = 0.106\,\mu\mt{m}$.}
    \label{fig:Pi_ON}
  \end{center}
\end{figure}

Thus it follows that
\begin{equation}\label{3.4}
\Pi^{\text{\tiny ON}}(t) \to \Pi_K(t)~~{\rm for}~~\gamma \to \infty,
~~\gamma^2/\Omega^2=~{\rm const}.
\end{equation}
In Fig.~\ref{fig:Pi_ON} it is shown how $\Pi_{\text{\tiny ON}}(t)$ approaches
$\Pi_K$ for large but finite $\gamma$.

\section{Limit of large $\Omega$ and deconvolution} \label{IV}
Experimentally, $\Omega$ is  easier to adjust than $\gamma$. Therefore
we also consider the limit of large $\Omega$, with $\gamma$ held
fixed. In this case one obtains
\begin{eqnarray}\label{4.1}
  \lambda_{\pm} &\simeq& \mp\frac{\Omega}{2} - \frac{i\gamma}{4}  \\
  q &=& \sqrt{k^2 +i m\gamma/\hbar},\qquad \Im\,q >0 \\
  k_{\pm} &\simeq& \sqrt{\pm\frac{m\Omega}{\hbar}} \pm \frac{1}{2}\Bigl(k^2 + 
  \frac{i m\gamma}{2\hbar}\Bigr)\sqrt{\frac{\pm\hbar}{m\Omega}}  \\
  R_1 &\simeq & -1 + (1-i)k\sqrt{\frac{\hbar}{m\Omega}}\\
  R_2 &\simeq& -(1+i)k\sqrt{\frac{\hbar}{m\Omega}},
\end{eqnarray}
to leading order in $\Omega$.
This yields
\begin{equation}\label{4.2}
  (1-|R_1(k)|^2)^{-\frac{1}{2}}\,(1-|R_1(k')|^2)^{-\frac{1}{2}} \simeq
  \frac{1}{2}\sqrt{\frac{m\Omega}{\hbar kk'}}
\end{equation}
and
\begin{multline}\label{4.3}
  \gamma \overline{\Phi^{(2)}_k(x)} \Phi^{(2)}_{k'}(x) \simeq
  \frac{\hbar\gamma}{2\pi m}\frac{kk'}{\Omega} \Bigl\{
          \Theta(-x)2\e^{i(\bar{q}-q')x} \\
          + \Theta(x) (-i\e^{-i \bar{k}_{+}x} - \e^{-i \bar{k}_{-}x})\\
          \times(i\e^{i k_{+}' x} - \e^{i k_{-}' x})\Bigr\}.
\end{multline}
When integrating over $x$, only the term $\e^{-i(\bar{k}_{+} -
  k_{+}')x}$ contributes in leading order of $\Omega$, and this gives
\begin{eqnarray} \label{eq:Pi_ON_lim}
 \Pi^{\text{\tiny ON}}(t) &\!\!\to&\!\! \frac{\hbar}{2\pi m} \int\! d kd
 k'\,\overline{\tilde{\psi}(k)} \tilde{\psi}(k')
 \e^{i\frac{\hbar}{2m}(k^2-k'^2)t}\nonumber\\&&\times \sqrt{kk'} \frac{\gamma}{\gamma +
   \frac{i\hbar}{m}(k^2-k'^2)}.
\end{eqnarray}
For $\gamma \to \infty$ one again obtains Kijowski's distribution, but
for finite $\gamma$ one has a delay in the arrival times. One can try
to eliminate this, as in Ref. \cite{Damborenea-PRA-2002}, by a
deconvolution with the first-photon distribution, $W(t)$, of an atom
at rest in the laser field, making the ansatz
\begin{equation}\label{4.4}
  \Pi^{\text{\tiny ON}}(t) = \Pi_{\mt{id}}(t) \ast W(t)
\end{equation}
for an ideal distribution $\Pi_{\mt{id}}(t)$. Clearly, $W(t)$ has the
meaning of an apparatus resolution function, similar to
Ref. \cite{Allcock}. In terms of Fourier transforms one
obtains from the ansatz
\begin{equation} \label{eq:conv_four}
\tilde{\Pi}_{\mt{id}}(\nu) =
\frac{\tilde{\Pi}^{\text{\tiny ON}}(\nu)}{\tilde{W}(\nu)},   
\end{equation}
where \cite{Kim-OC-1987}
\begin{equation}
  \frac{1}{\tilde{W}(\nu)} = 1 + \left(\frac{\gamma}{\Omega^2} +
    \frac{2}{\gamma} \right)i\nu + \frac{3}{\Omega^2}(i\nu)^2 +
  \frac{2}{\gamma\Omega^2}(i\nu)^3.
\end{equation}
From Eq. (\ref{eq:Pi_ON_fin}) one obtains
\begin{widetext}
\begin{eqnarray} \label{eq:Pi_ON_nu}
 \tilde{\Pi}^{\text{\tiny ON}}(\nu) &=& \gamma \int_{-\infty}^{\infty} d
 x~\int d kd k'~  
  \overline{\tilde{\psi}(k)} \tilde{\psi}(k')\,(1-|R_1(k)|^2)^{-1/2}
  \,(1-|R_1(k')|^2)^{-1/2} \nonumber\\
  &&\qquad\qquad\qquad\times\,\,2\pi\,\delta\left(\nu - \frac{\hbar}{2m}
    (k^2-k'^2)\right)
  \overline{\phi^{(2)}_k(x)} \phi^{(2)}_{k'}(x).
\end{eqnarray}
\end{widetext}
For large $\Omega$ one has $1/\tilde{W}(\nu) \simeq
1+2i\nu/\gamma$. Inserting this  into Eq.~(\ref{eq:conv_four}) and
using Eq.~(\ref{eq:Pi_ON_lim}) yields 
\begin{eqnarray}
 \tilde{\Pi}_{\mt{id}}(\nu) &=& \frac{\hbar}{2\pi m} \int d kd k'~  
  \overline{\tilde{\psi}(k)} \tilde{\psi}(k')
  \sqrt{kk'}\nonumber\\&&\times 
  2\pi\,\delta\left(\nu - \frac{\hbar}{2m} (k^2-k'^2)\right),   
\end{eqnarray}
and therefore, for any value of $\gamma$ and in the limit of strong driving, 
\begin{equation}
\Pi_{\mt{id}}(t) = \Pi_K(t).
\end{equation}

\begin{figure}[htbp]
  \begin{center}
    \includegraphics[width=7cm]{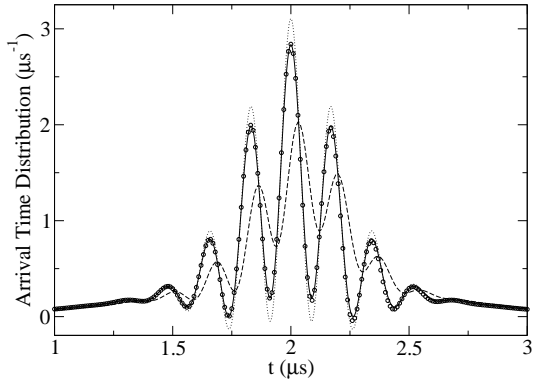}
    \caption{Excellent agreement between the deconvoluted operator-normalized
      distribution $\Pi_{\mt{id}}$ (white circles) and $\Pi_K$
      (solid line) for large $\Omega = 500\gamma$. Shown is also
      $\Pi^{\text{\tiny ON}}$ before deconvolution (dashed line). The initial
      wave packet is a coherent combination
      $\psi=2^{-1/2}(\psi_1+\psi_2)$ of two Gaussian states for the
      center-of-mass motion of a single cesium atom that become
      separately minimal uncertainty packets (with $\Delta x_1 =
      \Delta x_2 = 0.031\,\mu\mt{m}$, and average velocities $\langle
      v\rangle _1 = 18.96\,\mt{cm/s}$, $\langle
      v\rangle _2 = 5.42\,\mt{cm/s}$ at $x=0$ and
      $t=2\,\mu\mt{s}$). The flux (dotted) becomes negative in some place.}
          \label{fig:Pi_deconv}
  \end{center}
\end{figure}

The convergence of $\Pi_{\mt{id}}$ to Kijowski's distribution is shown
in Fig.~\ref{fig:Pi_deconv}. In this example the flux, which is a
limit of a deconvoluted fluorescence distribution without operator
normalization \cite{Damborenea-PRA-2002},
becomes negative.

\section{Connection with complex potentials}

The above approach to arrival times,
which was based on photon emissions, has
another interesting limit which establishes a connection with the 
complex-potential approach proposed by Allcock \cite{Allcock}. 
Consider now large $\gamma$ and $\Omega$, but with
$\Omega^2/\gamma=\mt{const}$ instead of $\Omega^2/\gamma^2$ as
before. Then a little 
calculation shows that in Eq.~(\ref{2.9}) the second
component $\psi^{(2)}_t
\sim \gamma^{-1/2}$ while the first component goes to
\begin{equation}\label{5.1}
  \psi^{(1)}(x,t) = \int_0^{\infty}d k\,\tilde{\psi}(k) \e^{-i\hbar k^2
    t/2m} \phi_k(x),
\end{equation}
where
\begin{eqnarray}\label{5.2}
  \phi_k(x) &=& \left\{
      \begin{array}{cl}
        \displaystyle \e^{i kx} + R \e^{-i
          kx},&\qquad x\leq 0\nonumber\\ 
        \displaystyle T \e^{i \kappa x},&\qquad x\geq 0
        \end{array}
        \right.\nonumber\\
  R &=& \frac{k-\kappa}{k+\kappa}\nonumber\\
  T &=& \frac{2k}{k+\kappa}\nonumber\\
  \kappa &=& \sqrt{k^2 + \frac{2i mV_0}{\hbar^2}},\qquad \Im\,\kappa
  > 0\nonumber\\
  V_0 &=& \frac{\hbar\Omega^2}{2\gamma}.
\end{eqnarray}
From Eq.~(\ref{5.1}) one obtains that $\psi^{(1)}_t$ satisfies the
one-dimensional Schrödinger equation
\begin{equation} \label{eq:1dimSG}
i\hbar\frac{d}{dt}{\psi}^{(1)}_t = (\hat{p}^2/2m -i
V_0\Theta(\hat{x})) \psi^{(1)}_t
\end{equation}
with the complex potential $-i V_0\Theta(\hat{x})$. Since
$\psi^{(2)}_t\to 0$ one has, from 
Eq.~(\ref{eq:Nt}),
\begin{equation}
  N(t) = ||\psi^{(1)}_t||^2
\end{equation}
and so, from $\Pi (t) = -d N/d t$ together with Eq.~(\ref{eq:1dimSG}),
\begin{equation} \label{eq:Pi1dim}
  \Pi(t) = \frac{2V_0}{\hbar} \int_0^{\infty}d x |\psi^{(1)}(x,t)|^2.
\end{equation}
This is consistent with Eq.~(\ref{eq:Pi2}) since $\gamma |\psi^{(2)}_t|^2$
remains finite.

Eqs.~(\ref{eq:1dimSG}) and (\ref{eq:Pi1dim}) provide a connection with the
complex-potential model of Allcock where the particle absorption rate
is taken as a measure for the arrival time. This model is here seen to
arise as a limiting case from the approach of
Ref.~\cite{Damborenea-PRA-2002}. It is also
obtained by considering, somewhat artificially, a position-dependent
Einstein coefficient, $\gamma(x) = \gamma\Theta(x)$, and using an
incoming state in the upper level, or from the 
irreversible detector model
put forward by 
Halliwell \cite{Halliwell}.   

The distribution in Eq.~(\ref{eq:Pi1dim}) is again not normalized to $1$,
and it is therefore natural to employ an operator normalization. With the
same arguments as in Section~III the operator-normalized distribution
is obtained as
\begin{eqnarray}
  \Pi^{\text{\tiny ON}}(t) &=& \frac{2V_0}{\hbar} \int_0^{\infty}d x \int d kd k'\,
  \overline{\tilde{\psi}(k)} 
  \tilde{\psi}(k')\nonumber\\
  &&\times (1-|R(k)|^2)^{-\frac{1}{2}}(1-|R(k')|^2)^{-\frac{1}{2}}\nonumber\\
  &&\times\,\, \overline{T(k)}T(k')\,
  \e^{i\hbar(k^2-k'^2)t/2m} \e^{-i(\overline{\kappa}-\kappa')x}.
  \nonumber \\&&
\end{eqnarray}
In the limit of strong interaction, $V_0\to\infty$, one again finds
that this goes to Kijowski's distribution,
\begin{equation}
  \Pi^{\text{\tiny ON}}(t) \to \Pi_K(t)\quad\mt{for}~V_0\to\infty.
\end{equation}
The advantage of the one-channel model is that it provides a simple
calculational tool for further, more complicated, arrival time
problems and that, by simple limits and operator normalization, it is
related to the  operational fluorescence approach as well as to
the axiomatic
distribution of Kijowski.

\section{Discussion}
In Ref.~\cite{Damborenea-PRA-2002} it had been pointed out that from
the algebraic structure of the arrival time distribution in the
operational fluorescence model it seemed impossible to obtain 
Kijowski's distribution by considering a suitable limit  since one
could not produce the necessary term $\sqrt{k}$. This term now
arises in the model through an operator normalization which
corresponds to the normalization approach of Ref.~\cite{Brunetti-PRA-2002}.
In simple, operational terms, this normalization can also be viewed as a
modification of the initial 
state in such a way that the detection losses, due in
particular to a strong 
laser driving,  are compensated. 
Our results provide a crucial step towards understanding
and clarifying the
physical content of Kijowski's 
distribution and, more precisely,  establish a set of operations and limits 
in which such a distribution could exactly be measured. In addition,
it has been shown in this paper that
Allcock's one-channel model, which was based on a somewhat ad hoc
complex absorption 
potential, is in fact  a limiting case of the fluorescence model and
also related to Kijowski's distribution through operator-normalization. 

Instead of the operator-normalized expression
of Eq.~(\ref{eq:Pi_ON_op}) one
could also consider the expectation value of  the not manifestly
positive expression 
$\hat{\Pi}_t^J \equiv \frac{1}{2}(\hat{B}^{-1}\hat{\Pi}_t +
\hat{\Pi}_t\hat{B}^{-1})$ whose time integral is also
1. Interestingly, in the limit $\gamma\to\infty$ and $\Omega^2/\gamma^2
=\mt{const}$ this yields for the distribution the quantum mechanical flux
$J$, discussed in Ref.~\cite{Damborenea-PRA-2002}.
 
In this paper we have concentrated on initial states with a
definite momentum sign, and freely moving particles. However, the
approach can be carried over to a more general setting and this will
be investigated elsewhere.

\appendix
\section{Explicit expressions for   $C_i(\alpha)$}

The constants $C_i(\alpha)$  in Eqs. (\ref{3.1}) and (\ref{3.3}) are
given by
\begin{widetext}
\begin{eqnarray}
  C_1 &=& \frac{2\sqrt{2}\alpha + (1+\alpha)^{3/2} -
    (1-\alpha)^{3/2}}{\sqrt{2}\alpha\sqrt{1-\alpha^2} +
    \sqrt{\alpha+1}(\alpha-1) +\sqrt{1-\alpha}(1+\alpha)} \\
  C_2 &=& \frac{2\sqrt{2}\sqrt{1-\alpha^2} \left(\sqrt{1+\alpha} -
      \sqrt{1-\alpha} \right)}{\sqrt{1+\alpha} (\sqrt{2}
    +\sqrt{1-\alpha}) (\alpha-1) + \sqrt{1-\alpha} (\sqrt{2}
    +\sqrt{1+\alpha}) (\alpha+1)} \\
  C_3 &=& \frac{1}{2}\left[ \sqrt{1+\alpha} (\sqrt{2}
    +\sqrt{1-\alpha}) (\alpha-1) +\sqrt{1-\alpha} (\sqrt{2}
    +\sqrt{1+\alpha}) (\alpha +1)\right]
\end{eqnarray}
\end{widetext}
with $\alpha \equiv \sqrt{1-4\Omega^2/\gamma^2}$.


\begin{thebibliography}{99}

          
\bibitem{ML00} J.G. Muga and C.R. Leavens, Phys. Rep. {\bf 338}, 353
(2000)

\bibitem{tqm} J.G. Muga, R. Sala and I.L. Egusquiza (eds.), 
{\it Time in Quantum Mechanics} (Springer, Berlin, 2002).

\bibitem{Allcock}
G.R. Allcock, Ann. Phys. (N.Y.) {\bf 53}, 253 (1969); {\bf 53}, 286
(1969); {\bf 53}, 311 (1969) 


\bibitem{Kijowski}
J. Kijowski, Rep. Math. Phys. {\bf 6}, 362 (1974) 

\bibitem{Werner} R. Werner, J. Math. Phys. {\bf 27}, 793 (1986). 

\bibitem{Yamada} N. Yamada and S. Takagi, Prog. Theor. Phys.  {\bf 85},
985 (1991); {\bf 86}, 599 (1991); {\bf 87}, 77 (1992).

\bibitem{Mielnik94}
{B.} {Mielnik},
{Found. Phys.} {\bf 24},
{1113} (1994). 

  

\bibitem{Leavens} C.R. Leavens, Phys. Rev. A {\bf 58},
840 (1998). 

\bibitem{BJ} P. Blanchard and A. Jadczyk, Helv. Phys. Acta {\bf 69}, 613
(1996). 

\bibitem{GRT} N. Grot, C. Rovelli, R. S. Tate, Phys. Rev. A {\bf 54}
4676 (1996).

\bibitem{Halliwell} J. J. Halliwell, Prog. 
Theor. Phys. {\bf 102} 707 (1999).

\bibitem{Finkelstein} J. Finkelstein, Phys. Rev. A {\bf 59}, 3218 (1999). 

\bibitem{Toller} M. Toller, Phys. Rev. A {\bf 59}, 960 (1999). 

\bibitem{KW} P. Kocha\'nski and K. W\'odkiewicz, Phys. Rev. A {\bf 60},
2689 (1999). 

\bibitem{AOPRU98}
Y.~Aharonov, J.~Oppenheim, S.~Popescu, B.~Reznik, and W.~G. Unruh,
Phys. Rev. A {\bf 57}, 4130 (1998). 

\bibitem{Aharonov-PR-1961}
Y. Aharonov, D. Bohm, Phys. Rev. {\bf 122}, 1649 (1961)

\bibitem{Leon-PRA-2000}
J. Le\'on, J. Julve, P. Pitanga, F.J. de Urr\'{\i}es, Phys. Rev. A
{\bf 61}, 062101 (2000)

\bibitem{BSPME00}
{A.~D.} {Baute},
{R.} Sala~Mayato,
{J.~P.} Palao,
{J.~G.} Muga, {and}
{I.~L.} Egusquiza,
{Phys. Rev. A} {\bf 61},
{022118} (2000). 

\bibitem{BEM01a}
{A.~D.} {Baute},
{I.~L.} {Egusquiza},
{J.~G.} Muga,
{Phys. Rev. A} {\bf 64},
{014101} (2001). 

\bibitem{Baute01}
A.~D. Baute,
I.~L. Egusquiza,
J.~G. Muga,
{Phys. Rev. A} {\bf{64}},
012501 (2001). 

\bibitem{DM} V. Delgado and J. G. Muga, Phys. Rev. A. {\bf 56}, 3425 (1997). 
\bibitem{AHN99} K. Aoki, A. Horikoshi,
and E. Nakamura, Phys. Rev. {\bf 62}, 022101 (2000).
\bibitem{Galapon} E. A. Galapon, Proc. Roy. Soc. {\bf 458}, 451 (2002). 
\bibitem{Wlodarz} J. J. Wlodarz, Phys. Rev. A {65}, 044103 (2002).
\bibitem{Leavens02}
C.~R. Leavens,
Phys. Lett. A {\bf{303}},
154 (2002).



\bibitem{EMNR03} I. L. Egusquiza, J. G. Muga, B. Navarro and A. Ruschhaupt, 
Phys. Lett. A, to appear. 

\bibitem{Damborenea-PRA-2002}
J.A.~Damborenea, I.L.~Egusquiza, G.C.~Hegerfeldt, J.G.~Muga,
Phys. Rev. A {\bf 66}, 052104 (2002).

\bibitem{Damborenea-JPB-2003}
J.A.~Damborenea, I.L.~Egusquiza, G.C.~Hegerfeldt, J.G.~Muga,
quant-ph/0302201

\bibitem{Giannitrapani-IJTP-1997}
R. Giannitrapani, Int. J. Theor. Phys. {\bf 36}, 1575 (1997)

\bibitem{Brunetti-PRA-2002}
R.~Brunetti, K.~Fredenhagen, Phys. Rev. A {\bf 66}, 044101 (2002)

\bibitem{Hegerfeldt93}
 G.~C. Hegerfeldt and T.~S. Wilser, in:
{\it Classical and Quantum Systems.} 
Proceedings of the Second International Wigner Symposium, July
1991, edited by H.~D. Doebner, W. Scherer, and F. Schroeck, (World
Scientific, Singapore, 1992), p. 104;
G.~C. Hegerfeldt,
\newblock Phys. Rev. A {\bf 47}, 449 (1993); G.~C. Hegerfeldt and
D.G. Sondermann, Quantum 
  Semiclass.~Opt.~{\bf 8}, 121 (1996). For a review cf.  M.~B. Plenio
  and P.~L. Knight, 
Rev. Mod. Phys. {\bf 70}, 101 (1998). The quantum jump approach is
essentially equivalent to the Monte-Carlo wavefunction approach of 
 J. Dalibard, Y. Castin and  K. M{\o}lmer,  
    Phys. Rev. Lett., {\bf68}, 580 (1992), and to the quantum trajectories of 
 H. Carmichael, {\em An Open Systems Approach to Quantum 
Optics}, Lecture Notes in Physics Vol.~18, (Springer, Berlin,  1993).

\bibitem{Kim-OC-1987}
M.S. Kim, P.L. Knight, K. Wodkiewicz, Opt. Comm. {\bf 62}, 385 (1987)



\end{thebibliography}
\end{document}